\documentclass[useAMS,usegraphicx]{mn2e}
\usepackage{rotate}
\usepackage{times}
\newif\ifAMStwofonts
\AMStwofontstrue

\def\gs{\mathrel{\hbox{\rlap{\hbox{\lower4pt\hbox{$\sim$}}}\hbox{$>$}}}}
\def\ls{\mathrel{\hbox{\rlap{\hbox{\lower4pt\hbox{$\sim$}}}\hbox{$<$}}}}

\def\Msun{M$_{\odot}$}

\def\xmm{{\it XMM-Newton}}

\def\hst{{\it HST}}

\def\asca{{\it ASCA}}

\def\xte{{\it RXTE}}
\def\xmm{{\it XMM-Newton}}

\def\swift{{\it Swift}}
\def\swiftgrb{{\it Swift Gamma Ray Burst Explorer}}

\def\et{{et al.\ }}

\def\erg{{\rm\thinspace erg}}

\def\keV{{\rm\thinspace keV}}

\def\Msun{\hbox{$\rm\thinspace M_{\odot}$}}

\def\s{{\rm\thinspace s}}

\def\ergps{\hbox{$\erg\s^{-1}\,$}}

\title[X-ray and optical variability of Seyfert 1s]
      {X-ray and optical variability of Seyfert 1 galaxies as observed with \xmm}

\author[R. Smith \& S. Vaughan]
       {R. Smith\thanks{E-mail: rjs47@star.le.ac.uk} and S. Vaughan \\
X-Ray and Observational Astronomy Group, University of Leicester, Leicester, LE1 7RH
}
\date{Accepted 2006 December 14; Received 2006 December 05; in original form 2006 October 03}
\pagerange{\pageref{firstpage}--\pageref{lastpage}}
\pubyear{2006}

\begin{document}
\maketitle
\label{firstpage}

\begin{abstract}
We have examined simultaneous X-ray and optical light curves of a sample of eight nearby Seyfert 1 galaxies observed using the EPIC X-ray cameras and Optical Monitor on board \xmm.
The observations span $\sim1$ day and revealed optical variability in four of the eight objects studied.
In all cases, the X-ray variability amplitude exceeded that of the optical both in fractional and absolute luminosity terms.
No clearly significant correlations were detected between wavebands using cross correlation analysis.
We conclude that, in three of the four objects in which optical variability was detected, reprocessing mechanisms between wavebands do not dominate either the optical or X-ray variability on the time-scales probed.
\end{abstract}



\section{Introduction}
\label{sect:intro}

\begin{table*}
\begin{center}
\caption{\label{tab:obs}Observation summary}
\begin{tabular}{ccccccccc}
\hline
\hline
Object          &Redshift     &On time   &Revolution & OM filter & No. OM exposures & p & OM rms & X-ray rms \\
&&(s)&&&&&\%&\% \\
(1)&(2)&(3)&(4)&(5)&(6)&(7)&(8)&(9)\\
\hline
1H 0707-495  &0.0411     &46018      &0159  &UVW2  &15    &0.067     &2.0  &40.1\\
Ark 120      &0.0322     &112130     &0679  &UVW2  &80    &0.051     &0.7  &2.0\\
Ark 564      &0.0246     &101774     &0930  &UVW1  &49    &0.088     &0.3  &22.8\\
MCG$-$6-30-15  &0.0077     &349286     &0301--0303 &U &275  &$<$0.01   &1.9  &28.2\\
Mrk 766      &0.0129     &129906     &0265  &UVW1  &75    &0.95      &0     &25.4\\
NGC 3783 (1) &0.0097     &40412      &0193  &UVW2  &30    &0.36      &0.4  &5.8\\  
NGC 3783 (2) &           &275633     &0371--0372 &UVW2 &53 &$<$0.01  &2.9  &20.7\\
NGC 4051 (1) &0.0023     &121958     &0263  &UVW1  &43    &$<$0.01   &1.5  &47.6\\
NGC 4051 (2) &           &51866      &0541  &UVW1  &15    &$<$0.01   &0.8  &43.8\\
NGC 7469     &0.0163     &164128     &0912--0913 &UVW2 &111 &$<$0.01 &1.9  &7.7\\
\hline
&&&&\\
\end{tabular}
\end{center}
\end{table*}

Active Galactic Nuclei (AGN) emit radiation over a broad range of wavelengths from radio through X-rays and are variable over this entire spectral range.
It is thought that the variability in different bands may be connected or driven by a single mechanism.
In this paper we examine the potential connection between optical and X-ray variability in Seyfert 1 galaxies.

Optical and UV continuum emission\footnote{Throughout this paper we use the term `optical' to refer to the optical-UV bandpass of the Optical Monitor} in AGN is believed to be thermal emission originating in the accretion disc surrounding the central supermassive black hole (SMBH).
The X-ray emission, which shows the most rapid variability, is thought to come from the hot `corona' close to the SMBH.
There are two main theories regarding a connection between these two bands: reprocessing of X-rays into thermal optical emission and Compton upscattering of optical photons to X-ray energies.
In the X-ray reprocessing scenario, X-rays from the corona are absorbed in the disc, heating it, which then re-radiates this energy as optical photons. 
The converse theory is that  photons from the disc enter the corona and are Compton upscattered by hot electrons to be emitted as X-ray photons.
It is currently unclear how important these mechanisms are in forming the continuum emission from AGN. 

If either of these processes are important in AGN then their effect should be observable in the variability of the source.
If optical photons drive the X-ray photons then an increase in flux in the optical light curve should be repeated in the X-ray band a short while later and vice versa.
By cross correlating X-ray and optical light curves to look for such a correlation, we can infer if and how the different emission regions are connected.
The traditional way to achieve this is to schedule simultaneous observations of a source on multiple telescopes.
\xmm{} provides a simpler way of performing these observations as it carries the Optical Monitor as well as three X-ray telescopes and can therefore perform X-ray and optical observations simultaneously.
Here we exploit this capability to examine a small sample of objects and study the relationship between X-ray and optical variability.

Previous studies of this subject have provided a confusing array of results.
In some cases, no correlation is detected on short time-scales but light curves are correlated when averaged over a longer period (e.g. NGC 4051; Peterson et al. 2000).
NGC 4051 was also studied by Shemmer et al. (2003) who found evidence that the X-rays were led by the optical emission by $\sim 2$ days.
For Ark 564, on the other hand, Shemmer et al. (2001) found that the UV lags the X-ray emission by $0.4$ days and optical lags UV by $\sim 2$ days. 
There are also cases where the X-ray and optical light curves are correlated with no lag.
In a 6 year observation of NGC 5548 by Uttley et al. (2003), a strong correlation between X-ray and optical light curves was detected at lag  $0 \pm 15$ days from data averaged at $30$ day intervals.
For further examples see Uttley (2005).
It is obvious that further work needs to be done to resolve the confusion surrounding reprocessing.

Only two previous studies have used the simultaneous observing capabilities of \xmm{} to examine variability correlations in AGN.
Mason et al. (2002) used a $1.5$ day long \xmm{} observation of NGC 4051 supplemented with data from \xte{} and found three prominent features in which the X-ray leads the optical emission. 
Using Monte Carlo simulations they derived a confidence of $85$ per cent in these features.
Ar\'{e}valo et al. (2005) analysed the long ($300$~ks) observation of MCG$-$6-30-15 and reported a significant correlation with the UV leading the X-rays by $1.8$ days. 
In this paper we re-examine these data along with data from six other Seyferts.

The remainder of this paper is organised as follows.
In section~\ref{sect:data} we describe the observations selected and the data reduction procedures. 
The robustness of our technique is discussed in section~\ref{sect:mc}.
Section~\ref{sect:results} contains the results of the analysis and in section~\ref{sect:disco} we compare our results to those made previously and discuss the implications of these results to the study of AGN variability.


\section{Observations and data reduction}
\label{sect:data}

\subsection{Sample definition and observations}
The satellite \xmm{} carries three X-ray telescopes and a co-aligned Optical Monitor (OM; Mason et al. 2001), comprising a $30$~cm Ritchey-Chr\'{e}tien telescope with a micro-channel plate intensified CCD. 
This gives a field of view of $17$ arcmin square and spatial resolution of $\sim 1$ arcsec, with $0.48$ arcsec pixels. 
The OM can take images using any of six narrow-band filters sensitive over the range $1800-6000$ \AA, thereby providing optical imaging and timing information simultaneous with the X-ray observations.
The combination of long X-ray exposures with simultaneous optical data makes \xmm{} ideal for probing the relation between X-ray and optical variability on short time-scales.

The \xmm{} Science Archive (XSA) contains all available data products from the satellite. The XSA was used to find long observations ($>30$~ks) of nearby (z $<0.1$) Seyfert 1 galaxies.
Objects were selected if there were $15$ or more exposures taken with a single filter of the OM. 
Eight objects were found to satisfy these criteria: 1H 0707-495, Ark 120, Ark 564, MCG$-$6-30-15, Mrk 766, NGC 3783, NGC 4051 and NGC 7469.
Four of these sources had observations in two or more revolutions in the same filter. 
A summary of the observations of these sources is given in Table ~\ref{tab:obs}.

\subsection{Predicted time delays for the sample}

The eight targets selected all have published black hole mass estimates, which allowed us to make simple predictions for the interband time delays that might be expected from reprocessing models.
Firstly we assumed all heating in the disc is generated by the viscosity of the disc itself.
Different radii of the disc are at different temperatures and can be modelled as blackbodies with a particular peak wavelength.
Using the effective wavelength of the appropriate OM filter, we calculated the temperature of a blackbody spectrum that peaks in the observed band using Wien's Law.
This temperature was then used with the standard formulae for a geometrically thin, optically thick accretion disc to determine the approximate radius from which the emission originates: 
\begin{equation}
T(r)\approx 6.3\times 10^{5}\left (\frac{\dot{M}}{\dot{M_{E}}}\right )^{1/4}M_{8}^{-1/4}\left (\frac{r}{R_{S}}\right )^{-3/4}
\end{equation}
where $\dot{M}$/$\dot{M_{E}}$ is the accretion rate as a fraction of the Eddington accretion rate, $M_{8}$ is the mass of the black hole in units of $10^{8}$\Msun{} and $r$/$R_{S}$ is the radius in terms of the Schwarzchild radius (Peterson 1997).
Using values of $0.01$ and $0.1$ for $\dot{M}$/$\dot{M_{E}}$, and published black hole masses (see Table ~\ref{tab:delay}), we derived a range for the radius of the optical region being observed. 
Assuming the X-ray emission comes from close to the central black hole, the light travel time to the radius of optical emission region gives a rough estimate for the time delay between these two regions.

As another check on the delay expected, we used the assumption that the disc is heated entirely externally by X-rays.
In this case, the energy of the blackbody peaking in the OM filter is comparable to the total X-ray luminosity. 
Using the temperature as calculated above, Stefan's Law was used to determine the radius of the optical emission region:
\begin{equation}
L_{X}=4\pi r^{2}\sigma T^{4}
\end{equation}
where $\sigma$ is Stefan's constant. 
The X-ray luminosity, L$_{X}$, is obtained from the $0.1-10$~\keV{} X-ray spectrum.
Again the radius was converted to an approximate time delay in light days.
The results of these calculations are shown in Table ~\ref{tab:delay}.
The majority of sources are expected to show time delays of less than a day which should theoretically be observable with \xmm.

\begin{table}
\begin{center}
\caption{\label{tab:delay}Estimated time delays}
\begin{tabular}{ccccc}
\hline
\hline
Object          &M$_{bh}$     &Int. delay   &L$_{x}$ & Ext. delay\\
&($10^{6}$\Msun)&(days)&($10^{44}$\ergps)&(days)\\
(1)&(2)&(3)&(4)&(5)\\
\hline
1H 0707-495  &2.3$^{1}$     &0.02--0.03      &0.42  &0.16\\
Ark 120      &150$\pm 19 ^{2}$     &0.25--0.54     &20.1  &1.10\\
Ark 564      &1.09$^{1}$     &0.01--0.03     &4.5  &0.98\\
MCG$-$6-30-15  &2.9$^{+1.8}_{-1.6}$$ ^{3}$     &0.03--0.07     &1.14 &0.69\\
Mrk 766      &0.63$^{4}$     &0.01--0.02  &1.83   &0.62\\
NGC 3783     &29.8$\pm 5.4 ^{2}$     &0.09--0.18      &2.55  &0.39\\
NGC 4051     &1.91$\pm 0.78 ^{2}$     &0.02--0.04     &0.03  &0.08\\
NGC 7469     &12.2$\pm 1.4 ^{2}$   &0.05--0.10     &3.9 &0.48\\
\hline
\end{tabular}
\begin{minipage}[t]{3.4in}
\small{$^{1}$ Zhou \& Wang (2005); $^{2}$ Peterson et al. (2004); $^{3}$ McHardy et al. (2005); \\
$^{4}$ Botte et al (2005)}\\
\small{{\sc Notes:} Values in column 3 are based on the assumption of internal disc heating by viscosity. The lower limits use a value of $\dot{M}$/$\dot{M_{E}}= 0.01$ and the upper limits use 0.1. X-ray luminosity (column 4) is for the band 0.1--10 keV. Values in column 5 are based on the assumption of external disc heating by X-rays.} 
\end{minipage}
\end{center}
\end{table}

\subsection{Optical light curves}
The Observation Data Files (ODFs) for each source were extracted from the XSA and processed using the task {\tt omichain} of the \xmm{} Science Analysis System (SAS v6.5.0).
{\tt Omichain} takes the ODFs and performs all standard OM processing steps on them to produce the final products - images and source lists.
Once the images have been created, {\tt omichain} performs source detection and photometry routines.
Finally the source positions are converted from detector to sky coordinates and a combined source list from all exposures is created.

To create an optical light curve, photometry must be performed on each exposure of each object. 
Although {\tt omichain} automatically performs photometry on each image, the presence of host galaxy emission in some images causes the routine to flip between  `extended' and `point source' modes of photometry, and so does not perform a uniform extraction of all the images.
As the aim is to obtain accurate photometry of the nucleus and not include the host emission this is obviously undesirable.
In order to ensure uniform data reduction we used custom made scripts to perform aperture photometry and apply the appropriate OM corrections to the resulting data, modelled closely on the SAS processing tasks {\tt omdetect} and {\tt ommag}.

Simple aperture photometry was applied to the images produced by {\tt omichain} (after the modulo-8 correction has been applied by the task {\tt ommodmap}). 
Based on the user defined source position (in detector coordinates), the exact source location was obtained by repeatedly centroiding on the counts within a small box, then the counts within a circular aperture were accumulated. 
For all sources, except MCG$-$6-30-15, a circle of radius $12$ pixels was used as the source aperture to match the calibration (see below).
In the case of MCG$-$6-30-15, a $12$ pixel radius aperture overlapped a nearby star so, to avoid this, an aperture of $7$ pixels was used for all exposures of MCG$-$6-30-15.
For most sources, an annulus placed around the source aperture was used to extract the background count rate.
The background annulus was placed at least $30$ pixels out from the source to avoid the extended PSF of the UV filters.
In the cases of Ark 564, MCG$-$6-30-15 and NGC 4051 the host galaxy is visible around the AGN. 
As the host galaxy should not be varying in flux and it is the flux variability, not absolute flux, that we are studying, we chose to attempt no subtraction of the host galaxy flux from the flux of the central AGN. 
To get an accurate value for the actual background of the field, a circular background aperture away from the host galaxy was used in these cases. 

In order to produce accurate count rates from aperture photometry of OM images there are five corrections that must be applied - point spread function (PSF1), coincidence loss (CL), deadtime (DT), a UV specific PSF correction (PSF2) and time-dependent sensitivity degradation (TDS). 
Coincidence loss is a loss of flux that occurs when more than one event is collected in the same pixel in a single CCD frame, and the CL correction is a non-linear correction that has been calibrated empirically and found to be self-consistent when applied to an aperture radius of $6$ arcsec ($12$ pixels).
Before the CL correction can be applied, the PSF1 correction must therefore be implemented to scale the count rate from the aperture used to an effective aperture radius of $12$ pixels.
When the CCD is read out there is a short amount of time, the frame transfer time, when any incoming photons are not detected. 
The deadtime factor corrects for this loss.
In the UV filters the wings of the PSF are more extended than in the optical filters so a further correction is made to count rates from exposures in these filters to account for the loss of count rate from the wings.
Finally the time-dependent degradation factor takes account of the fact that the performance of the OM gradually decreases over time. 
The uncertainties on each flux point were derived by propagating the Poissonian error on the source and background count rates.

\subsection{X-ray light curves}
X-ray light curves were produced using data from the European Photon Imaging Camera pn camera (EPIC-pn) onboard \xmm.
Event lists for each observation were extracted using the standard processing procedure, {\tt epchain}.
In order to check for background flaring, $10-15$~keV light curves were extracted from off-source regions and examined by eye for periods of high, variable background. 
Where such intervals occurred they were removed by filtering out all time intervals during which the count rate in the high energy light curve exceeded a critical value that was chosen to be slightly above the quiescent level, but was slightly different for each observation.
After filtering out background flares, source light curves were extracted from a $40$ arcsec circular region using only single or double-pixel events in the range $0.2-10.0$~\keV, and background subtracted using a nearby source-free region.
The light curves were binned such that the X-ray sampling is comparable to the sampling of the OM light curves.

\begin{figure*}
\begin{center}
\scalebox{0.5}{\rotatebox{270}{\includegraphics{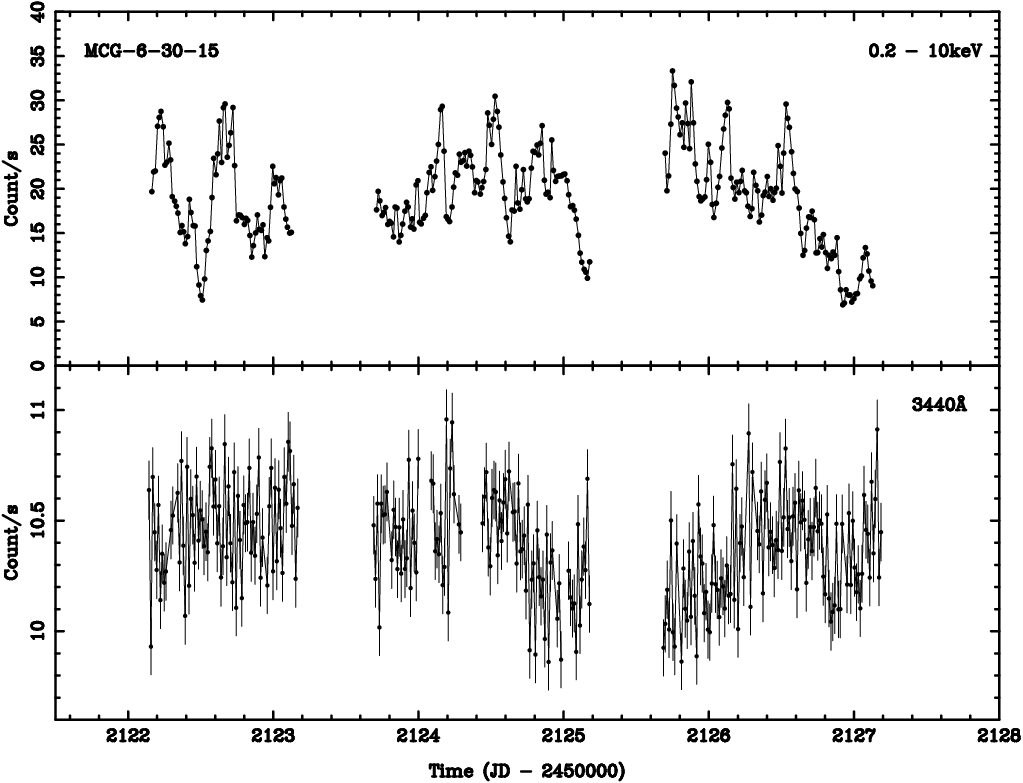}}}
\scalebox{0.5}{\rotatebox{270}{\includegraphics{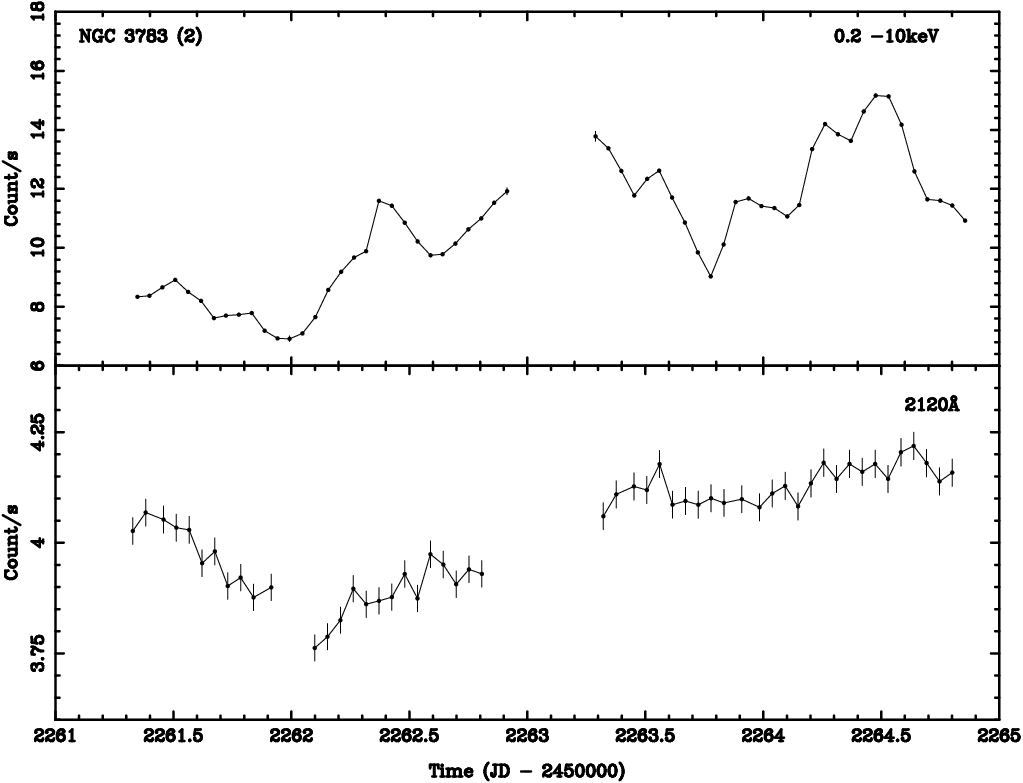}}}
\scalebox{0.5}{\rotatebox{270}{\includegraphics{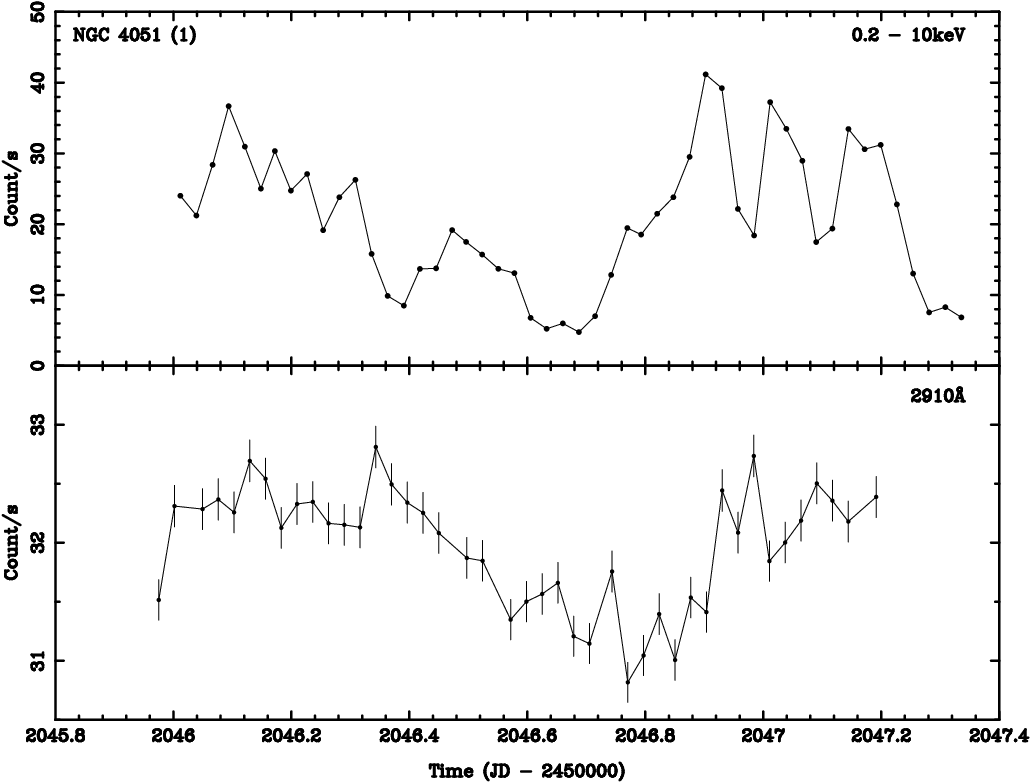}}}
\scalebox{0.5}{\rotatebox{270}{\includegraphics{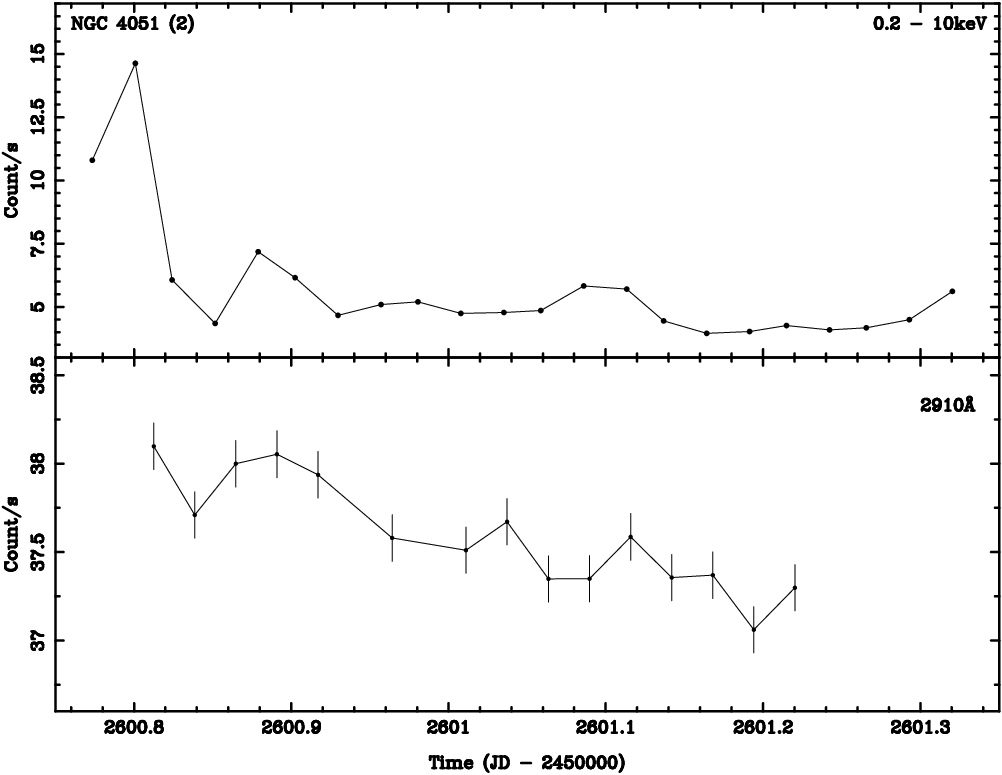}}}
\scalebox{0.5}{\rotatebox{270}{\includegraphics{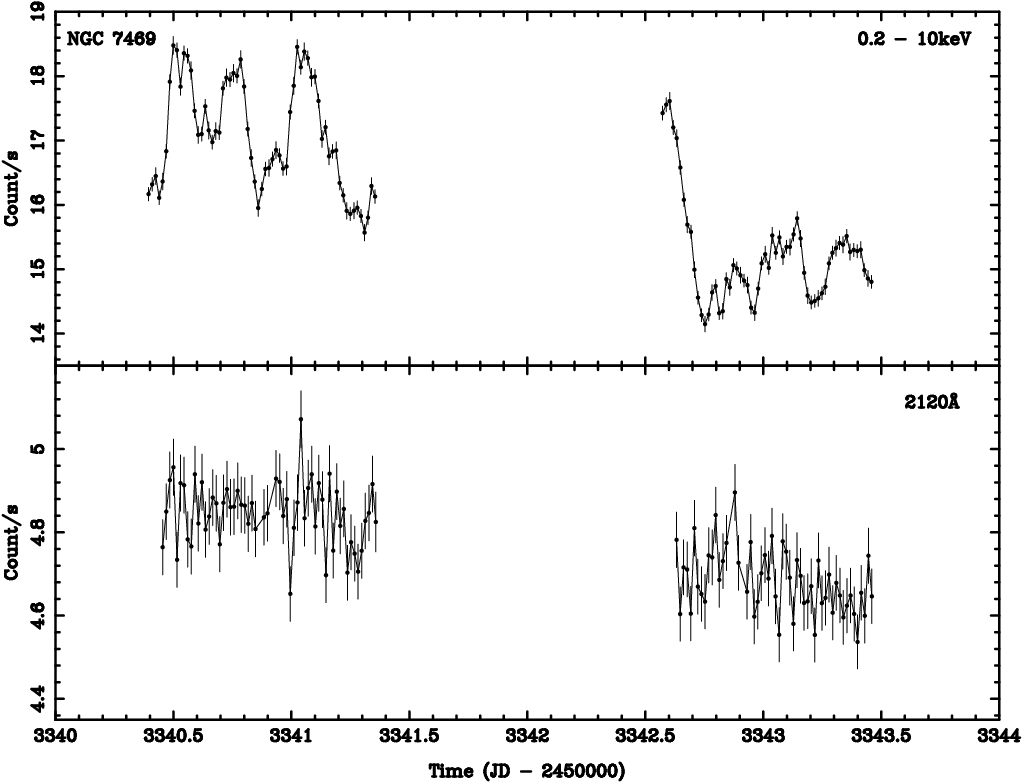}}}
\end{center}
\caption{
X-ray ($0.2-10$ keV) and optical light curves of MCG$-$6-30-15 (Rev. 0301--0303), NGC 3783 (Rev. 0371--0372), NGC 4051 (Rev. 0263), NGC 4051 (Rev. 0541) and NGC 7469 (Rev. 0912--0913). X-ray light curves are binned to match the optical sampling. Error bars are included for all count rate measurements, but for several of the X-ray light curves the error bars are smaller than the data point symbols.
}
\label{fig:lightcurves}
\end{figure*}

Figure~\ref{fig:lightcurves} shows the optical/UV and X-ray light curve for the sources with detectable X-ray and optical variability.
MCG$-$6-30-15, NGC 3783 and NGC 7469 all had consecutive observations which have been combined.

\section{Testing the method}
\label{sect:mc}

\subsection{Monte Carlo tests}

In order to test the robustness of our light curve extraction (outlined above) to PSF losses and host galaxy contamination, random data were simulated and subjected to the same aperture photometry procedure. 
As the spacecraft pointing can change during an observation by as much as a few arcsec, causing the target source to drift several pixels across the image, this effect was also included in the simulations. 
The simulations were designed to match the MCG$-$6-30-15 data, chosen because it is the longest dataset and shows the most contamination from its host galaxy. 
The simulation procedure was as follows.

The `true' source position on a $224 \times 224$ pixel detector grid was defined as $(x_0,y_0)$ based on the average detector position of MCG$-$6-30-15 in the real OM images. 
The effect of random pointing changes was simulated by giving a random offset to the precise position of the point source in each image. 
A point source was then added to the image at the new position $(x_0+\Delta x,y_0 + \Delta y)$ using the PSF model obtained from the latest calibration files. 
In practice a differential PSF model was calculated from the cumulative PSF model listed in the calibration files, and linearly interpolated to produce a function sampled with $0.1$ pixel resolution out to $12$ pixels. 
This was used to define the image of the point source from a smooth PSF projected onto a relatively coarse pixel grid. 
The host galaxy emission was modelled as an ellipsoid with an exponential surface density profile, with spatial parameters and normalisation (relative to the point source) taken from the spatial analysis of Ar\'{e}valo \et (2005), kindly supplied by Dr Ar\'{e}valo. 
The source image was then normalised to match the typical counts per image from the observations, and a typical (flat) background level ($\sim 3.3$ ct/pix) was added. 
The counts per pixel were then randomly drawn from a Poisson distribution to simulate random counting statistics. 
The simulated images were then subjected to aperture photometry.
Specifically, repeated centroiding was used to find the exact source position in each image, then counts were accumulated from a circular region and the PSF1 correction was applied exactly as for the real data. 
The effects of the other corrections, namely CL, TDS and DT, were not included.
The latter two corrections are simply constant factors determined for each exposure, and the CL is well calibrated for aperture photometry (assuming the PSF1 correlation is correct).

The simulations were designed to reproduce the effects of aperture photometry on a constant point source, the position of which is slightly different for each image, superposed on a realistic host galaxy image and flat constant background level, all subjected to Poisson variations. 
The end result of $200$ simulations was a time series that could be fitted with a constant to test for spurious variability.  
Using random position offsets $(\Delta x, \Delta y)$ drawn from a Gaussian of mean zero and standard deviation $0.2$ pixels, the result was a fit statistic of $\chi^2 = 211.29$ with $199$ degrees of freedom (dof), a perfectly acceptable fit ($p = 0.262$), indicating no statistically significant variability was present given $200$ images. 
A longer simulation using $5000$ images did show significant variability ($\chi^2 = 5432.56$ for $4999$ dof, $p = 1.2 \times 10^{-5}$) but only at the $0.4$ per cent level, which is too small to be detected in less than $200$ images. 
This excess variability is also far smaller than the measured amplitudes in any of the observations (see below).

Allowing for a $10$ pixel drift in both $x$ and $y$ directions over the course of $200$ images, comparable to the true drift during the MCG$-$6-30-15 observation, plus a random offset for each image, gave similar results: $\chi^2 = 205.19$ for $199$ dof ($p = 0.39$). 
There was no statistically significant correlation ($r = 0.044, ~ p = 0.53$) between the measured flux and the offset between the centroid and the stationary `first guess' position $(x_0,y_0)$ given to the photometry algorithm, indicating that the photometry method is not biased by even fairly large changes in spacecraft pointing. 
The same results were obtained when the host galaxy was not included in the simulations, proving that contamination by host galaxy light within the source aperture does not affect the variability. 
Changing the random offsets $(\Delta x,\Delta y)$ to have a standard deviation of $0.5$ pixels in each direction also gave entirely consistent results.

\subsection{Empirical tests using field stars}

\begin{table}
\begin{center}
\caption{\label{tab:fs}Field sources used for testing}
\begin{tabular}{cccc}
\hline
\hline
Field target     &RA     &Dec   &$p$\\
&(degrees)&(degrees)&\\
(1)&(2)&(3)&(4)\\
\hline
Ark 564      &340.593    &29.725     &0.02\\
             &340.701    &29.756     &0.70\\
MCG$-$6-30-15  &203.966    &-34.137    &0.21\\
Mrk 766      &184.611    &29.813     &0.80\\
NGC 4051     &180.786    &44.539     &0.21\\
             &180.793    &44.541     &0.56\\
             &180.763    &44.541     &0.77\\

\hline
\end{tabular}
\end{center}
\end{table}

Satisfied that our method performed well using simulated data we then tested it on actual data from the OM.
For several of the sources in the sample (Ark 564, Mrk 766, MCG$-$6-30-15), stars were visible in the field of view.
To prove that the method does not introduce spurious variability, the process was tested on a selection of these field stars.
The field around Ark 564 contains many objects.
For our field star tests we chose two objects of similar brightness to Ark 564 which were far enough removed from other objects in the field to allow accurate light curves to be extracted.
The other fields examined here contained few objects other than the target so all field stars (unless contaminated by being too close to the target) were tested. 

Table~\ref{tab:fs} summarises the results of these tests.
A $p$ value of $\gg0.01$ indicates that a light curve is non-variable. 
Three of the four stars tested proved to be non-variable proving that the method used here does not introduce spurious variability.
One of the stars tested in the Ark 564 field has a $p$ value of $0.02$ implying possible variability in the source.
We performed CCF analysis between this object and the Ark 564 light curve to ensure the detected variability is not a systematic problem also affecting Ark 564. 
No correlations were found between the two objects.
We suggest that the star is intrinsically variable and does not cause a problem with our method.

As another test of whether or not the method is robust, light curves for extended sources in the NGC 4051 host galaxy were extracted.
These will not vary on the time-scales probed here so a non-varying light curve should definitely be produced.
The light curves extracted did indeed prove to be non-varying confirming that the method does not introduce spurious variability. 


\section{Results}
\label{sect:results}

To determine whether or not the optical light curve is variable, a $\chi^2 $ test was performed on each light curve.
The resulting $p$ values are given in Table~\ref{tab:obs}.
Values of $p < 0.01$ indicate variation at the $99$ per cent confidence level or greater and values between $0.01$ and $0.05$ are possible detections of variability ($95 - 99$ per cent confidence). 
Variability was detected in the OM light curves at 99 per cent confidence in four out of eight objects.

Also shown in Table~\ref{tab:obs} is the optical and X-ray rms variability amplitude for each source. 
The typical error in these values is around one per cent so, depending on the brightness of the source, an object with rms amplitude greater than one should be detectable as varying.
The optical rms values shown here are probably underestimated as they include the total flux of the source with no host galaxy subtraction.
For the four sources that show significant variability in both optical and X-ray bands, the rms variability amplitude was calculated in luminosity units, as tabulated in Table~\ref{tab:rms}.

\begin{table}
\begin{center}
\caption{\label{tab:rms}RMS variability}
\begin{tabular}{ccc}
\hline
\hline
Object          &OM rms    &X-ray rms\\
&($10^{42}$\ergps)&($10^{42}$\ergps)\\
(1)&(2)&(3)\\
\hline
MCG$-$6-30-15  &0.2  &35\\
NGC 3783 (2) &3.3  &51\\
NGC 4051 (1) &0.08 &2.9\\
NGC 4051 (2) &0.5  &0.7\\
NGC 7469     &7.3  &30\\
\hline
&&\\
\end{tabular}
\end{center}
\end{table}

A standard tool for testing and measuring correlations between two time series is the cross correlation function (CCF), which gives the degree of correlation between the two series as one series is shifted backwards and forwards in time relative to the other. 
Unfortunately, when the data are not exactly evenly sampled one cannot apply the standard, `textbook' formulae for estimating the CCF (Box \& Jenkins 1976; Priestley 1981; Brockwell \& Davis 1996).
There are two reasons why the present data are not evenly sampled: the OM images are rather evenly spaced in time but there are a few gaps, and observations spanning more than one \xmm\ revolution will be split by the orbital gap. 
Suggested solutions to the problem of uneven sampling include interpolating the data onto an even grid, or calculating the correlation using pairs of data points that have a lag falling within a finite range (i.e. $\tau_{ij} = t_x(i) - t_y(j)$ falls in the interval $[\tau, \tau + \Delta \tau$]). 
Specific algorithms that have been used in AGN monitoring include the interpolated cross correlation function (ICCF; Gaskell \& Peterson 1987), the discrete correlation function (DCF; Edelson \& Krolik 1988) and the z-transformed discrete correlation function (ZDCF; Alexander 1997).  
All three of these were used in the present analysis and the results were found to be consistent.

A second problem that plagues CCF estimation from AGN data is that the data are often non-stationary, meaning that the mean and/or variance of a short light curve will change over time, due to variations on time-scales longer than the length of the observation. 
Following the advice of Welsh (1999) two precautions were used to mitigate the side-effects of non-stationarity.
Firstly a linear function was fitted to the data (using ordinary least-squares) and removed to suppress any long time-scale variation, and secondly the mean and variance were calculated `locally', using only the data contributing to a particular lag. 
Following another suggestion in Welsh (1999) the correlation coefficient was computed using both the raw data and the ranked data (each data point being correlated was replaced with its rank). 
In principal using the ranked data should improve sensitivity to non-linear correlations and make the statistic more robust to outliers, compared to using the raw data. 
In practice the two were found to be in close agreement.

Confidence limits on the CCF were computed using {\it Bartlett's formula} (Bartlett 1955) as an approximate but simple test of the significance of any correlations. 
In particular, standard deviation of the distribution of CCF points around an expected value of zero was computed using equation 11.1.9 of Box \& Jenkins (1976). 
This assumes that in reality the two series are not correlated, and can be used to test against this hypothesis. 
Under the assumption of no cross-correlation (i.e. the null hypothesis that the expectation value of the CCF is zero) Bartlett's formula may be written as:

 \begin{equation}
 \sigma^2\{ {\rm CCF}(\tau_j) \} = \frac{1}{N_{\rm pair}}\sum_{i} {\rm ACF_1}(\tau_i) {\rm ACF_2}(\tau_i)
 \end{equation}

where the two ${\rm ACF}$ functions are the empirical auto-correlation functions and the summation is carried out over {\it all} measured lags.
Here $N_{\rm pair}$ is the number of pairs of data points contribution at the lag $\tau_{j}$.

The 95 per cent confidence region was calculated as $\pm 1.96 \sigma$. 
The confidence region for the correlation coefficient at a given lag scales with $N_{\rm pair}^{-1/2}$. 
The bounds are smallest at small time delays, where the two time series overlap the most, and increase towards the edges of the CCF. 
The CCFs were examined over a range of lags spanning, in either direction, half the observation duration. 
At longer lags the number of pairs that contribute to the CCF calculation rapidly decreases, leading to highly uncertain CCF estimates.
It should be noted however that Bartlett's formula is strictly only valid when the number of data points $N$ is large and the data are stationary. 
As the X-ray/optical data do not satisfy either condition well, the confidence limits should be considered as rough estimates of the size of the true confidence region. 
Bartlett's formula is discussed in most standard references for time series analysis including: Jenkins \& Watts (1968; sect 8.2.1), Box \& Jenkins (1976; sect 11.1.3), Priestley (1981; sect 9.5), Bendat \& Piersol (1986; sect 8.4), Kendall \& Ord (1990; sect 11.3), or Brockwell \& Davis (1996; sect 7.3.4).

Previous papers discussing inter-band correlations in AGN variability have made use of Monte Carlo simulations to test the significance of any possible correlations. 
In principle the Monte Carlo approach should be robust but in practice it does rely on the power spectra of the two light curves being well understood in order to produce realistic simulations. 
Technically the CCF is not a `pivotal' statistic, meaning it is sensitive to `uninteresting' parameters such as the power spectral shape, and so simulations must be made as carefully as possible. 
For the present analysis it was not possible to obtain reliable power spectra from the limited OM data, and so Bartlett's formula (which uses the empirical auto-correlation function) was used as a guide of significance. 
It is also worth noting the results of Monte Carlo tests are sensitive to which of the pair of light curves are randomised -- whether the X-ray, or optical, or both dataset are randomised can substantially affect the apparent significance, and as yet there is no accepted preference.
This choice is not required for Bartlett's formula to be used.

\begin{figure*}
\begin{center}

\scalebox{0.3}{\rotatebox{270}{\includegraphics{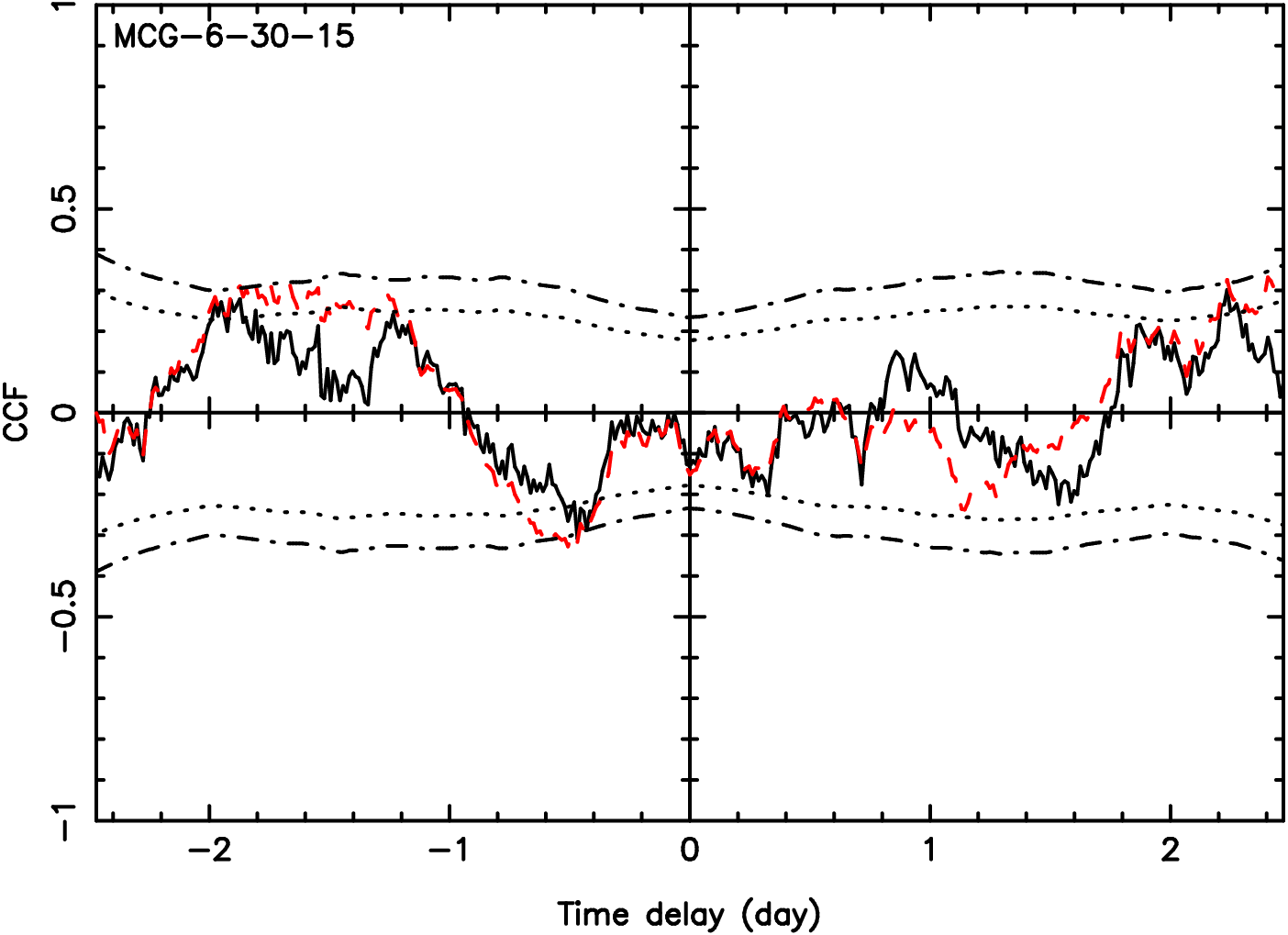}}}
\scalebox{0.3}{\rotatebox{270}{\includegraphics{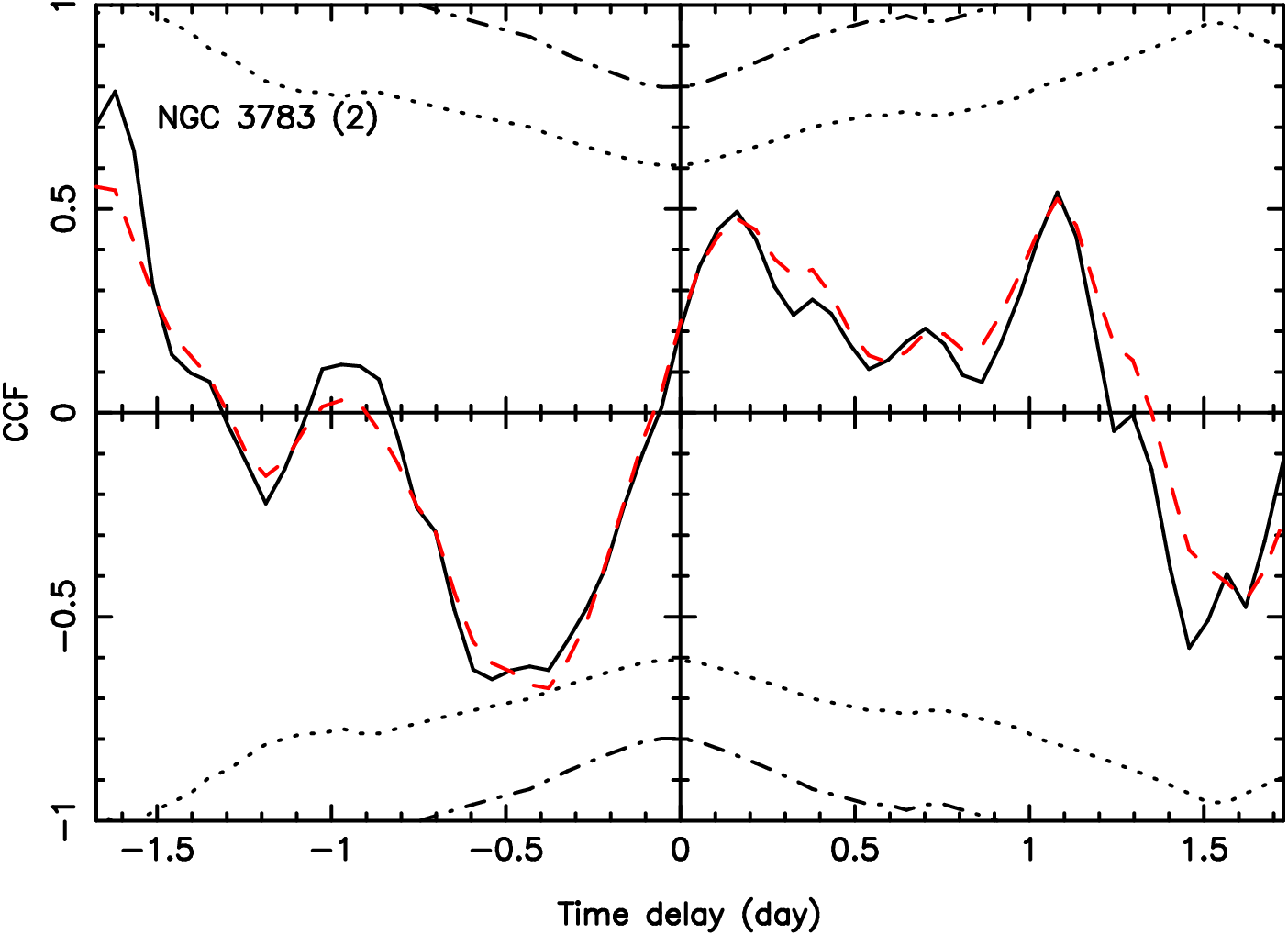}}}
\scalebox{0.3}{\rotatebox{270}{\includegraphics{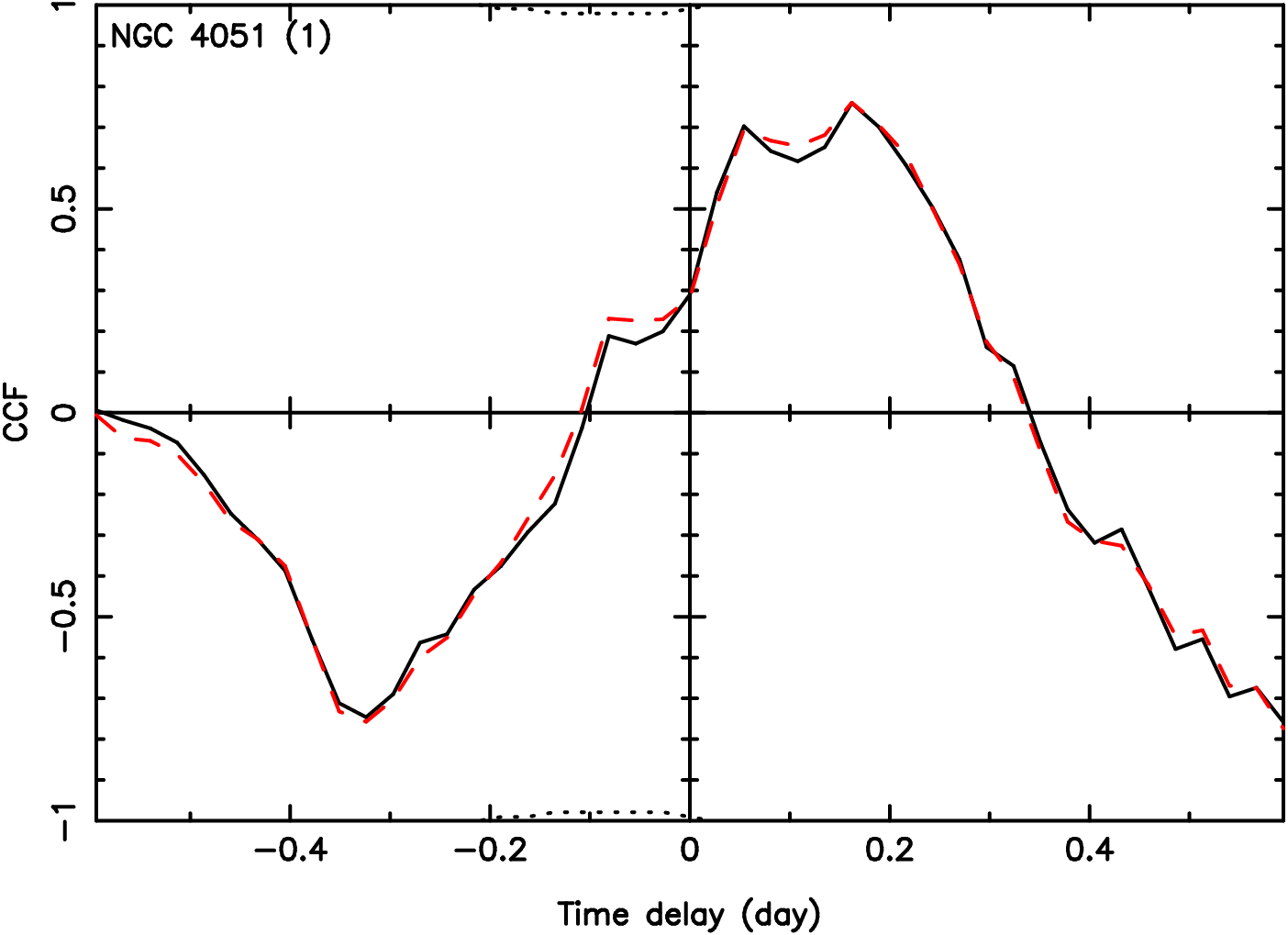}}}
\scalebox{0.3}{\rotatebox{270}{\includegraphics{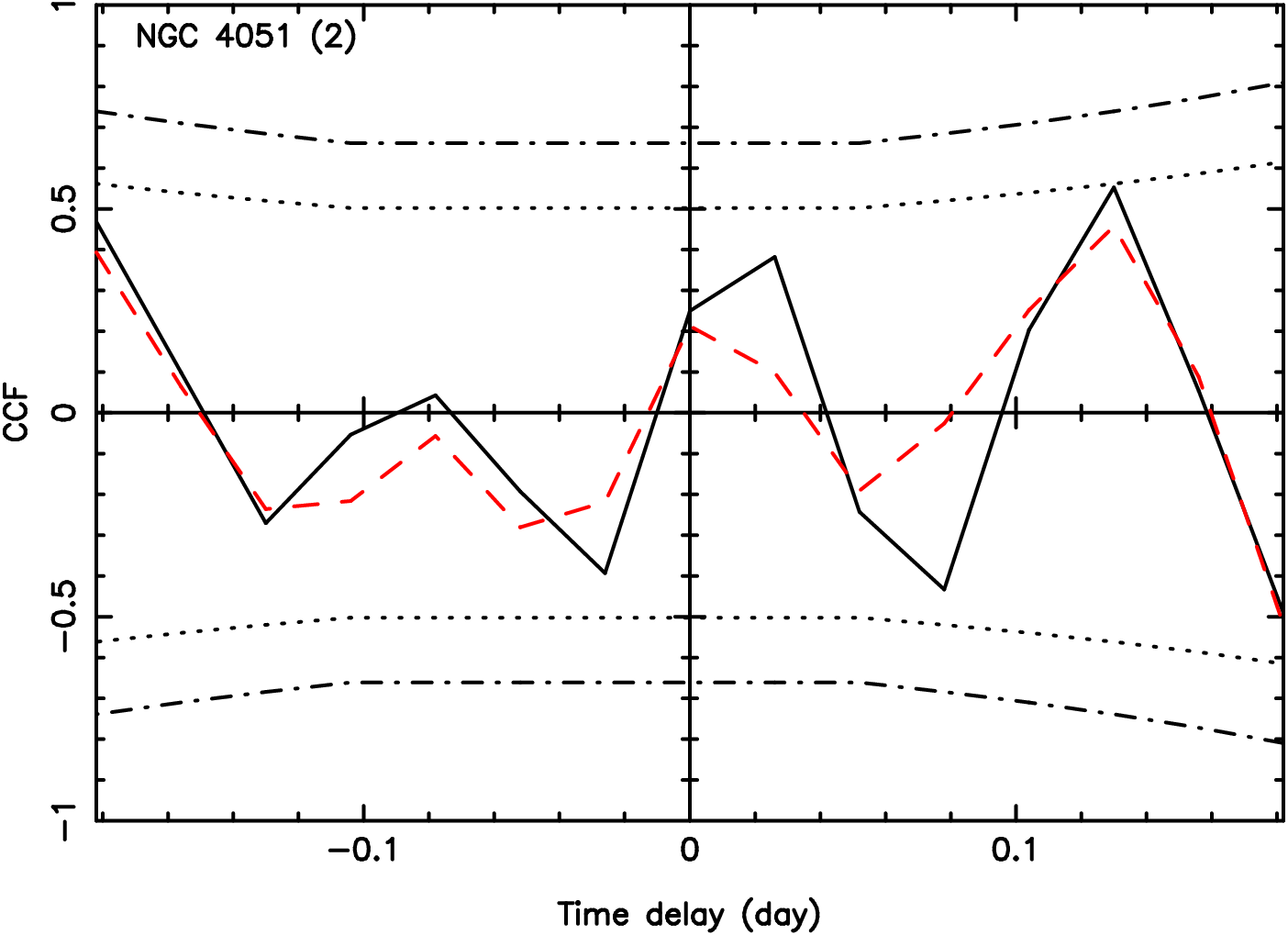}}}
\scalebox{0.3}{\rotatebox{270}{\includegraphics{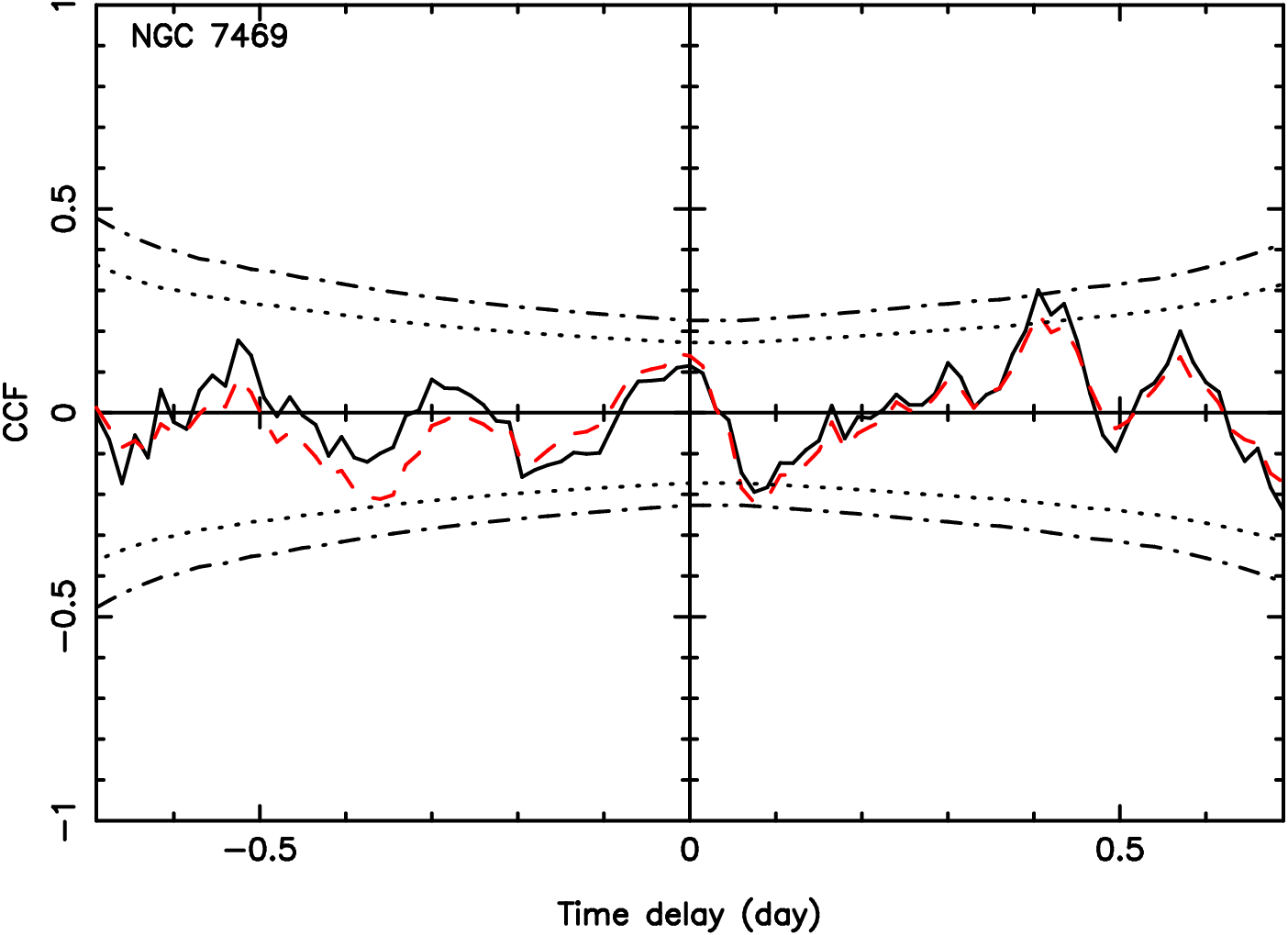}}}
\end{center}
\caption{
CCF plots of MCG$-$6-30-15 (Rev. 0301--0303) NGC 3783 (Rev. 0371--0372), NGC 4051 (Rev. 0263 - 95 per cent confidence limit just visible), NGC 4051 (Rev. 0541) and NGC 7469 (Rev. 0912--0913). The DCF is plotted in black and the ICCF is the red dashed line. The dotted and dot-dashed lines represent the 95 and 99 per cent confidence limits respectively. 
}
\label{fig:ccfs}
\end{figure*}

Figure ~\ref{fig:ccfs} shows the CCF plots for the objects studied which displayed optical and X-ray variability. 
A positive lag in the plots implies the X-ray variations lead those in the optical and vice versa for a negative lag.
The 95 and 99 per cent confidence limits are indicated by dotted and dot-dashed lines respectively.
In the case of NGC 4051 (Rev. 263) and NGC 3783 the $95$ per cent confidence region is sufficiently broad that these data do not impose useful constraints on the possibility of a correlation.
The error bounds, as computed using the Bartlett formula, were large in these cases because the empirical ACFs for both optical and X-ray time series were very broad.
Particularly in the case of NGC 3783 the X-ray power spectrum over the time-scales studied here is rather steep, leading to smooth light curves and a broad ACF, and hence a large error in the Bartlett formula. 
This accounts for the fact that realisations of processes with steeper power spectra (or equivalently, broader ACFs) are more `weakly non-stationary'.
However, in the cases of NGC 4051 (Rev. 541), NGC 7469 and MCG$-$6-30-15 the CCF is sufficently well constrained to exclude a strong correlation, i.e. $95$ per cent confidence limit of $|{\rm CCF}(\tau)| < 0.5$, over a plausible range of $\tau$.

Some previous studies (e.g. Nandra et al. 2000) have detected a correlation between the optical light curve and X-ray photon index but no correlation between optical and X-ray light curves. 
In light of this we also cross correlated the optical light curves of the objects studied here with the $0.7-2/2-10$~\keV{} hardness ratio (which should track the photon index) but found the cross correlation functions to be very similar to the ones between X-ray and optical flux.

\section{Discussion}
\label{sect:disco}

\subsection{Summary of results}
\label{sect:sum}
Using data from the OM and X-ray telescopes aboard \xmm{} we have examined a sample of eight objects to look for evidence of reprocessing between wavebands.
Optical variability on short time-scales was detected at $99$ per cent confidence in four out of eight objects and rms variability amplitudes in all cases show that the X-ray variability is stronger, in both fractional terms and absolute luminosity units, than the optical variability.
In three out of the four objects showing detectable optical variability, the cross correlation function is sufficiently well constrained to exclude a strong X-ray/optical correlation and hence reprocessing on time-scales of $<1$ day is not a dominant mechanism.

\subsection{Comparison with previous studies}
\label{sect:past}
\subsubsection{Observations using \xmm{} data}
The observation of NGC 4051 in Revolution 0263 was also investigated by Mason et al. (2002) for inter-band correlations.
They supplemented the \xmm{} data with X-ray observations from \xte{} and their CCF analysis covers positive lags up to two days.
Over the region that their CCF plot overlaps with ours ($0-0.6$ days) the results are consistent.
We observe a peak in the CCF around $0.2$ days with a value of $0.75$ similar to their peak of $\sim 0.6$.
Using simulations of uncorrelated red-noise light curves they place a confidence of $85$ per cent in the correlation. 
Our peak lies between $68.3$ and $95$ per cent confidence levels, certainly not significant at $99$ per cent.
However, as mentioned in section~\ref{sect:results}, the confidence regions in this case are not well enough constrained to comment on the correlation.

The OM observation of MCG$-$6-30-15 was previously analysed by Ar\'{e}valo et al. (2005), who reported a positive correlation with DCF$_{max} = 0.82$ at $160$ ks (around $1.8$ days) with the UV leading the X-ray data. 
Using Monte Carlo simulations they estimated a confidence limit of $98.5$ per cent (i.e. a $p$-value of $0.015$). 
By contrast our analysis found a much lower CCF peak ($0.35$) which was not considered a significant detection of correlation.

There are several differences between the analysis of Ar\'{e}valo et al. (2005) and that of the present paper. 
In terms of the light curve extraction, Ar\'{e}valo et al. (2005) performed photometry on attitude corrected images by fitting an empirical PSF model to the source image, while simultaneously modelling the host galaxy emission.
The present analysis used aperture photometry on (detector space) images based on the SAS processing pipeline tasks, including corrections for coincidence loss, deadtime and the counts/frame dependence of the PSF model, none of which were included by Ar\'{e}valo et al. (2005).
These differences in extraction technique explain the apparent difference between the two reductions of the same data.
Ar\'{e}valo et al. (2005) also used coarser binning of the light curves, but applying the same binning to our light curves did not substantially decrease the discrepancy.
As shown in Section~\ref{sect:mc} the aperture photometry method employed in this paper is robust to host galaxy presence and to pointing changes but, unlike the analysis of Ar\'{e}valo et al. (2005), our method includes all the known time-dependent flux corrections that should be applied to OM data.
Therefore, in principle at least, this reduction should be the more reliable.

In practice most of the difference between the correlation strengths is due to the more conservative approach employed in the present analysis for calculating the CCF. 
Ar\'{e}valo et al. (2005) presented the DCF after normalising by the noise-subtracted variances (i.e. the variances used in the denominator of the correlation coefficient formula had the mean square error subtracted). 
The present analysis used ICCF, DCF and ZDCF methods without the noise-subtraction term, employing linear and rank-order correlation coefficients, the latter should be more robust to outliers.
Including the noise-subtraction term increased the CCF value by $\approx 0.2$ at the claimed lag, but this only affects the absolute value of the CCF and not its significance (since the effect should be calibrated out during the significance test).

A second difference is that, in the present analysis the time series were `detrended' before calculating the CCF by subtracting off a linear function.
As recommended by Welsh (1999) this helps remove any bias caused by very long time-scale variations manifested as quasi-linear trends over the data that can lead to spuriously high correlation coefficients. 
Applying the detrending reduced the CCF by $\approx 0.2$ at the claimed lag. 
If the correlation at $\sim 1.8$-d was robust it should not be affected by removal of longer time-scale ($\sim 5$-d) variations.

\subsubsection{Observations from other telescopes}
Several other sources discussed here have also been studied previously: Ark 564, NGC 4051 and NGC 7469.

Ark 564 was observed by Papadakis et al. (2000) over a period of $21$ days at the Skinakas Observatory in optical and simultaneously in X-rays with \asca.
Papadakis et al. (2000) probed slightly longer time-scales than we do but, in line with the trend reported here, they find no significant correlations between the two bands.
Shemmer et al. (2001) examined Ark 564 over a longer period ($50$ days) in UV with \hst{} and X-ray with \xte{} and \asca. 
They found evidence of UV emission following X-ray by $\sim 0.4$ days.

Shemmer et al. (2003) studied NGC 4051 and found evidence for a correlation between X-ray and optical bands with a lag of less than one day. 
Their campaign covered a period of three months with approximately daily observations and so probed a longer time-scale than we examine here.
NGC 4051 was also investigated by Peterson et al. (2000). 
On short time-scales they found no evidence for correlations. 
Having smoothed the data in $30$ day bins however a correlation consistent with zero lag is detected.  

A $30$ day observation of NGC 7469 by Nandra et al. (1998) found that at least two variability mechanisms are required to describe their result.
Cross correlation of UV and X-ray light curves gave evidence for a lag of around four days.
This lag explained the differing times of maxima in the two light curves but failed to describe how the minima in the light curves occurred simulataneously.

As none of the studies described above probe the same short time-scales examined here it is difficult to make a comparison between them and the results described in this paper.
An overall picture seems to be emerging of no correlation between X-ray and optical emission over sub-day periods but significant correlations only appearing on much longer time-scales.

\subsection{Implications for AGN variability}
In line with other work we find that X-ray rms variability exceeds that in the optical in all cases.
However, in the objects with well constrained CCF (MCG$-$6-30-15, NGC 4051 (Rev. 0541) and NGC 7469), the absence of a strong correlation implies the majority of the variance in the optical band on short time-scales is uncorrelated with the short time-scale X-ray variations.
Papadakis et al. (2000) suggest that this behaviour could be explained if only a small section of the accretion disc is visible to the X-rays or if the X-ray emission is anisotropic.
Two of the observations showing optical variability, NGC 3783 and NGC 4051 (Rev. 0263) have poorly constrained CCF confidence regions and so a firm conclusion on these objects can not be drawn. 

In terms of the overall picture emerging from variability studies this solidifies earlier findings (Edelson et al. 2000; Uttley et al. 2003) that, on short time-scales, X-ray and optical variations in Seyfert 1 galaxies are uncorrelated. 
Many papers (see section~\ref{sect:past}) find some evidence for correlations on longer time-scales.
It seems that rapid X-ray and optical variations in Seyfert 1 galaxies occur through separate mechanisms but long-term variations are somehow connected.

Intensive observations with \xmm{} are not suitable for probing longer time-scale variations.
However, one promising alternative is the \swiftgrb.
\swift{} carries co-aligned X-ray (XRT; Burrows et al. 2005) and UV/optical (UVOT; Roming et al. 2005) telescopes covering essentially the same bandpass as \xmm, but is in a low-Earth orbit making it better suited to regular `snapshot' monitoring of sources over weeks-months.
The combination of \xmm{} and \swift{} is therefore capable of probing X-ray/optical correlations with unprecedented detail on all time-scales from hours to weeks.


\section*{Acknowledgments}

We thank the anonymous referee for useful comments and suggestions.
RJS and SV acknowledge PPARC for financial support.
This paper is based on observations obtained with \xmm, an ESA science mission with
instruments and contributions directly funded by ESA Member States and
the USA (NASA).  
This research has made use of the NASA/IPAC Extragalactic Database
(NED) which is operated by the Jet Propulsion Laboratory, California
Institute of Technology, under contract with the National Aeronautics
and Space Administration.


\bsp
\label{lastpage}

\end{document}